# Achieving Different Stoichiometries and Morphologies in Vapor Phase Deposition of Inorganic Halide Perovskites: Single or Dual Precursor Sources?

Tomáš Musálek[1,2], Petr Liška[1,2], Amedeo Morsa[3], Jon Ander Arregi[2], Matouš Kratochvíl[4], Dmitry Sergeev[3,5], Michael Müller[3], Tomáš Šikola[1,2], Miroslav Kolíbal[1,2,*]

[1]Institute of Physical Engineering, Brno University of Technology, Technická 2, 616 69 Brno, Czechia

[2]CEITEC BUT, Brno University of Technology, Technická 10, 61669 Brno, Czechia

[3]Institute of Energy Materials and Devices, Structure and Properties of Materials (IMD-1), Forschungszentrum Jülich, 52425 Jülich, Germany

[4]Materials Research Centre, Faculty of Chemistry, Brno University of Technology, Purkyňova 118, 612 00 Brno, Czech Republic

[5]NETZSCH-Gerätebau GmbH, Selb, D-95100, Germany

* mail: kolibal.m@fme.vutbr.cz

**Abstract**

Inorganic halide perovskites have become attractive for many optoelectronic applications due to their outstanding properties. While chemical synthesis techniques have been successful in producing high-quality perovskite crystals, scaling up to wafer-scale thin films remains challenging. Vapor deposition methods, particularly physical vapor deposition and chemical vapor deposition, have emerged as potential solutions for large-scale thin film fabrication. However, the control of phase purity during deposition remains problematic. Here, we investigate single-source ($CsPbBr_3$) and dual-source ($CsBr$ and $PbBr_2$) vapor deposition techniques to achieve phase-pure $CsPbBr_3$ thin films. Utilizing Knudsen Effusion Mass Spectrometry, we demonstrate that while the single-source $CsPbBr_3$ evaporation is partially congruent, it leads to compositional changes in the evaporant over time. The dual-source evaporation, with a precise control of the $PbBr_2/CsBr$ flux ratio, can improve phase purity, particularly at elevated substrate temperatures at excess PbBr$_2$ conditions. Our results give direct evidence that the growth is CsBr-limited. Overall, our findings provide critical insights into the vapor phase deposition processes, highlighting the importance of evaporation conditions in achieving the desired inorganic perovskite stoichiometry and morphology.

I. **Introduction**

Inorganic halide perovskites have received a tremendous and still increasing attention of the scientific community because of their excellent optoelectronic properties, qualifying them for use in solar cells, light-emitting diodes[1], X-ray detectors, etc.[2]. This rapid burst of scientific studies was fuelled mainly by chemical synthesis preparation techniques, which provide high-quality perovskite (nano)crystals at very low cost[3]. However, upscaling this approach to wafer-scale thin films poses a great challenge. Recently, vapor deposition techniques have come to focus; evaporation from a solid source under vacuum conditions is an established technology e.g. in semiconductor lasers, allowing the preparation of thin films with atomic-scale precision[4]. Building on large know-how from different material systems, deposition from a vapor phase has the potential to become a dominant technology for a large-scale deposition of inorganic halide perovskite thin films as well.

The perovskite preparation technique is vital to determine the relevant properties of the resultant thin film[5]. Many experimental studies on vapor deposition of halide perovskites have been recently summarized by timely and important reviews[6-8]. Importantly, if effusion cells are utilized, vapor deposition offers ultimate control of the evaporation fluxes, thus allowing to engineer, e.g. grain size[9,10] or intermixing of different chalcogenides[11]. The efficacy of vapor deposition has soon been recognized and demonstrated by the deposition of the perovskite solar cells[12] and the perovskite light emitting diodes[13,14]. However, many issues prevail. Deposition of CsPbBr$_3$ by evaporation of precursors of different composition results in thin films of CsPbBr$_3$ mixed with unreacted CsBr and PbBr$_2$ or other phases of a different stoichiometry. Hence, the phase-purity control is poor even for the most prominent inorganic halide perovskite, CsPbBr$_3$. Overall, the literature reviews give a clear picture of current know-how in the field: a comprehensive understanding of the processes involved in every stage of the growth (evaporation, transport, and growth) is lacking[6-8].

The most prominent approaches to vacuum-based vapor deposition are chemical vapor deposition (CVD) and evaporation from Knudsen-like cells, the latter belonging to the family of physical vapor deposition (PVD) techniques. Usually, CVD relies on tubular quartz reactors, where, for instance, the precursor powders are placed at a certain position within a temperature gradient across the furnace. The vaporized precursors are dragged by a carrier gas toward the substrate, which is located at different positions within the tube furnace; here, the vapors condense into a thin layer. A sealed variant of CVD (without the carrier gas) is called the chemical vapor transport (CVT); here, the sample size is limited. CVD and CVT reactors that utilize solid precursors are single-use; after each growth run, the precursors need to be refilled. In contrast, PVD techniques are typically housed in larger high-vacuum apparatuses, where the Knudsen cells do not require refilling after every deposition, thus ideally providing stable and reproducible growth runs. However, typically in perovskites growth, this is difficult to achieve, as it is shown in this paper.

Most often, the precursors in Knudsen cells are located in resistively heated crucibles. Another possibility, specifically for organic-inorganic perovskites, is the pulsed laser deposition (PLD)[15] or its variant, resonant infrared matrix-assisted pulsed laser evaporation (RIR-MAPLE)[16]. These technologies seem very promising but remain unexplored.

Regardless of the technology in use, the deposition of a multicomponent materials is done from individual precursors of each component placed separately in different heating zones[17] or separate evaporation cells[18-20], as well as from a single precursor[11,21,22]. The former approach promises a good stoichiometry control of the resulting material, where most studies focused on finding the optimum flux ratio. However, the necessary premise for the deposition of phase-pure materials from multiple separated precursors is that the stoichiometry of the vapor phase (i.e. the flux ratio) is transferred to the grown material despite many processes happening on the substrate surface during growth. For instance, depositing two components at a specific molar ratio does not always lead to stoichiometry preservation of the deposit, as a high mobility at the surface is a key factor for the molecules to reach low-energy positions. It is worth noting that one ideally needs a high mobility of both components simultaneously[23], which is usually not the case (despite the fact that, for example, the $CsPbBr_3$ formation from gaseous CsBr and $PbBr_2$ is exothermic, − 386 kJ/mol[13], providing an additional thermal energy to deposited molecules).

In order to identify the „turning knobs" that can be used to tune the proces, we can learn from established III-V semiconductor deposition processes, where the sample temperature controls the surface concentration of one of the components, promoting in this way the formation of a phase-pure material with an ideal stoichiometry. This is the case, e.g. of GaAs[24], and SnSe[25], where the growth is Ga- and Sn-limited, respectively. The use of high growth temperatures ensures that the excess As (or Se) is desorbed, resulting in an optimum crystal growth. Such an approach has not been well studied for halide perovskites, as the knowledge of surface processes that occur during growth from multiple precursor fluxes is rather poor. The general disadvantage of this approach is that the fluxes should be low so that near-equilibrium structures can form, thus limiting the growth rate. Specifically, for halide perovskites, a possible proposed solution to this issue was a sequential evaporation of each component followed by annealing of the whole multilayer stack. However, the annealing is another critical step, facing several kinetic restrictions; perfect mixing is ensured only if the temperature is high enough to promote interdiffusion while avoiding a PbBr$_2$ desorption. A solution could be the deposition of multiple layer sequences, but in principle the difficulties with annealing prevail[26]. Additionally, out of vacuum, issues arise with different humidity conditions[27].

Evaporation from a single source has the potential to overcome these issues under the specific condition, i.e. that the evaporation of a compound occurs congruently. That means the composition of the vapor is reflecting that of the evaporant[28,29]. This is possible if the precursor is in the form of a nanoscale powder or if the precursor pellet fragments into nanoscale parts and decomposes only after fragmentation. The latter process has recently been documented to occur during the melting of CsPbBr$_3$[30]. However, the congruent evaporation of CsPbBr$_3$ has not yet been validated, despite speculations in the literature[31]. Understanding the single source evaporation of CsPbBr$_3$ is complicated as different stoichiometries of the resulting growth products are reported, being Cs or Pb rich[32,33]. Post-growth alloying strategies have been employed to increase the purity of the CsPbBr$_3$ phase[32-34], but this approach faces similar issues as the aforementioned annealing of multilayer stacks[26,27]. An apparent solution is the deposition at elevated temperature. Although this increases complexity, at the same time it introduces possibility of tuning the resulting morphology, from polycrystalline layers at lower temperatures to nanowires at higher temperatures[17,22], although contradictory results have been reported so far[11].

Here, we study a single source (CsPbBr$_3$)- and dual source (CsBr and PbBr$_2$)-deposition strategies with the objective of depositing phase-pure CsPbBr$_3$. We analyse two critical phases of growth by separately analysing the evaporation products using Knudsen Effusion Mass Spectrometry (KEMS) and the growth products at different sample temperatures by relevant analytical techniques. We show that the single source decomposition of CsPbBr$_3$ is partially congruent; however, the precursor composition changes during the evaporation. It is observed that this problem can be resolved to some extent by raising the sample temperature, which ensures the phase purity of the deposit by desorbing excess $PbBr_2$. Similarly, the dual-source evaporation yields a high-purity $CsPbBr_3$ evaporant at elevated sample temperatures, because growth is CsBr-limited. At different sample temperatures, the compounds of other stoichiometries, namely $Cs_2PbBr_5$ and $Cs_4PbBr_6$, are detected on the samples, either pure or mixed with $CsPbBr_3$, CsBr, or $PbBr_2$.

II.  **Methods**:

Perovskite $CsPbBr_3$ precursor synthesis: $PbBr_2$ powder (Sigma Aldrich, 99.999% purity) was dissolved in 48% aqueous HBr, and CsBr powder (Sigma Aldrich, 99.999% purity) was dissolved in $H_2O$. In both cases. the molar ratio of the dissolved powders was 1:1. Next, the solutions are mixed, resulting in the precipitation of an orange solid, which was suction filtered, washed with ethanol, and dried under vacuum. For co-evaporation experiments, pure $PbBr_2$ and CsBr powders were used.

Perovskite evaporation: All evaporation experiments were conducted within an high vacuum (HV) chamber, maintaining a base pressure of $1\times10^{-8}$ mbar. The chamber was equipped with two custom-built effusion cells, both aligned at an incident angle of 30° relative to the substrate. During the operation, the pressure typically stabilized in an order of $10^{-7}$ mbar. The sample holder was fitted with a calibrated pyrolytic boron nitride (pBN) heater to ensure a precise temperature control. Si(111) substrates were used in all the experiments.

Grazing-incidence X-ray diffraction (GIXRD): Structural analysis and phase identification of the deposited Cs-Pb-Br samples was performed via in-plane GIXRD measurements using a Rigaku Smartlab (9 kW) diffractometer with Cu-Kα radiation (λ = 1.54 Å) and an incident parallel beam setting. A 5° in-plane Soller slit was utilized in both the incident and diffracted optics. The X-ray source and the detector were set to a grazing angle (ω = 0.8º, 2θ = 1.6º) and intensity profiles were acquired by scanning the 2θ/χ angle, where the detector arm is laterally scanned in order to look for crystalline planes that are perpendicular to the sample surface. The grazing-incidence measurement geometry maximizes the signal arising from the thin film in comparison to that of the substrate. The recorded peaks were compared to crytallographic databases (see Fig. S1).

Scanning electron microscopy (SEM) was performed using ThermoFisher Verios 450L and FEI Versa microscopes.

X-Ray Photoelectron Spectroscopy (XPS): compositional analysis was performed using a Kratos Axis Supra XPS instrument utilizing the monochromated Al Kα radiation (1486.6 eV) and a hemispherical analyzer set in a high magnification mode with a pass energy of 20 eV. The electron emission angle was set along the normal to the surface. All the spectra were acquired with an energy step of 0.05 eV and integrated by utilizing several sweeps. The spectra are not shifted. The intensity and fluence of an X-Ray source was determined according to Ref. [35].

Photoluminescence and Raman spectroscopy analysis: spectra were obtained using a confocal microscope set-up Witec Alpha 300R with a built-in spectral camera and continual laser illumination. The wavelength of the laser light used for excitation was 532 nm and the optical power of the laser varied from 0.5 mW to 10 mW. The spectra were normalized with respect to the intensity of Si Raman peak (520 cm$^{-1}$). All the measurements were done using an objective with 100× magnification, working distance 0.3 mm and numerical aperture 0.9. The diffraction grid for photoluminescence measurements and detailed Raman spectra observations was 600 g/mm and 1200 g/mm, respectively.

Knudsen effusion mass spectrometry experiments were conducted using a FINNIGAN MAT 271 spectrometer at Forschungszentrum Jülich. Prior to sample introduction, the Knudsen cells were preconditioned by heating at 1000°C for 12 hours to remove impurities. For each measurement, approximately 40 mg of sample powder was loaded into the cell. Ionization was performed using an electron beam generated by a tungsten cathode, operating at an energy of 60–70 eV and an emission current of 0.47 mA. Two types of measurement series were carried out: isothermal measurements, where vapor species were analyzed at a constant temperature over time, and polythermal measurements, which were taken at different temperature intervals. Full experimental details are provided in the Supporting Information.

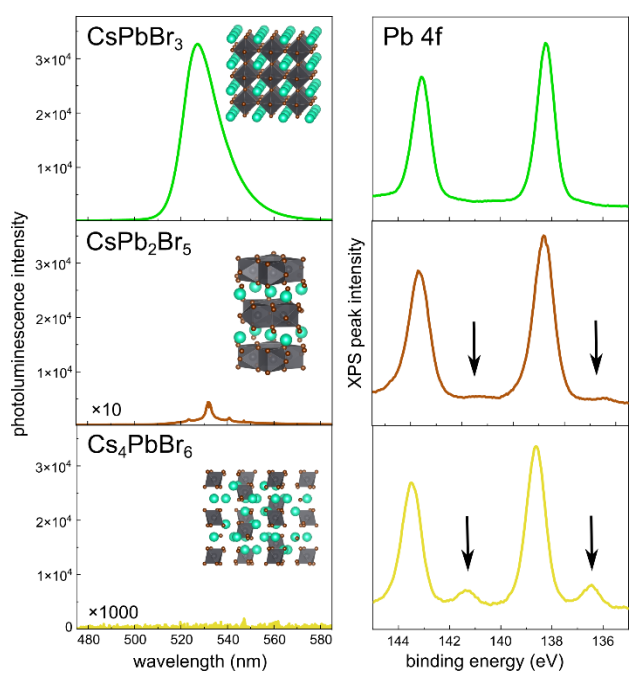

Fig. 1: Evaporation from different sources (dual source, CsBr+PbBr$_2$ or single source, CsPbBr$_3$) at different temperatures gives rise to inorganic Cs-Pb-Br-based perovskites of different stoichiometries: CsPbBr$_3$, as well as CsPb$_2$Br$_5$ and Cs$_4$PbBr$_6$. These perovskites exhibit strikingly different properties. The left panel shows photoluminescence intensity; by far the most prominent light emission comes from CsPbBr$_3$, while the other perovskites are almost non-emissive. In the right panel, a resistance against X-ray-induced damage is documented by XPS spectra after exposure to 1486 eV X-Rays from Al K$\alpha$ source with a fluence of $5\times10^{16}$ photons/cm$^2$. A reduction of higher valence states of Pb to metallic Pb$^0$ is documented by the appearance of a component marked with arrows.

### III. Results

The strategy of growing Cs-Pb-Br-based inorganic perovskites by the vapor phase deposition builds on the possibility of various species coming from the evaporator(s), thus allowing specific perovskite self-assembly on the surface, directed by the molar ratios of individual components. First, we demonstrate the dissimilar optical properties for the compounds with different stoichiometries, which possess distinct crystal structures (see insets in Fig. 1, left panel). Typically, strong photoluminescence (PL) is observed for direct band-gap $CsPbBr_3$, while PL is significantly quenched for $CsPb_2Br_5$ and $Cs_4PbBr_6$ due to an indirect band-gap in their electronic structure. The absence of PL for $Cs_4PbBr_6$ and $CsPb_2Br_5$ is well known and documented in the literature[36,37]. The previously established controversy about PL properties of the compounds with dissimilar stoichiometries has been resolved by growing pure crystals without $CsPbBr_3$ inclusions, as these inclusions were responsible for the PL signals previously seen for $Cs_4PbBr_6$ and $CsPb_2Br_5$. Therefore, the PL data in Fig.1 show that it is also possible to grow pure crystals from the vapor phase. $CsPbBr_3$ exhibits a very intense PL and is at the same time the most resistant compound to degradation upon exposure to X-rays. In Fig. 1 (right panel), we show that under an X-ray fluence of $5\times10^{16}$ photons/cm$^2$ (1486 eV X-Rays), the $Pb^0$ component (metallic Pb) arises in the XPS spectrum for $CsPb_2Br_5$ and $Cs_4PbBr_6$, while it is absent for $CsPbBr_3$. The $Pb^0$ component has previously been used as a reliable indicator of a Pb reduction[38], which subsequently results in a collapse of the perovskite structure and degradation of the deposited layer.

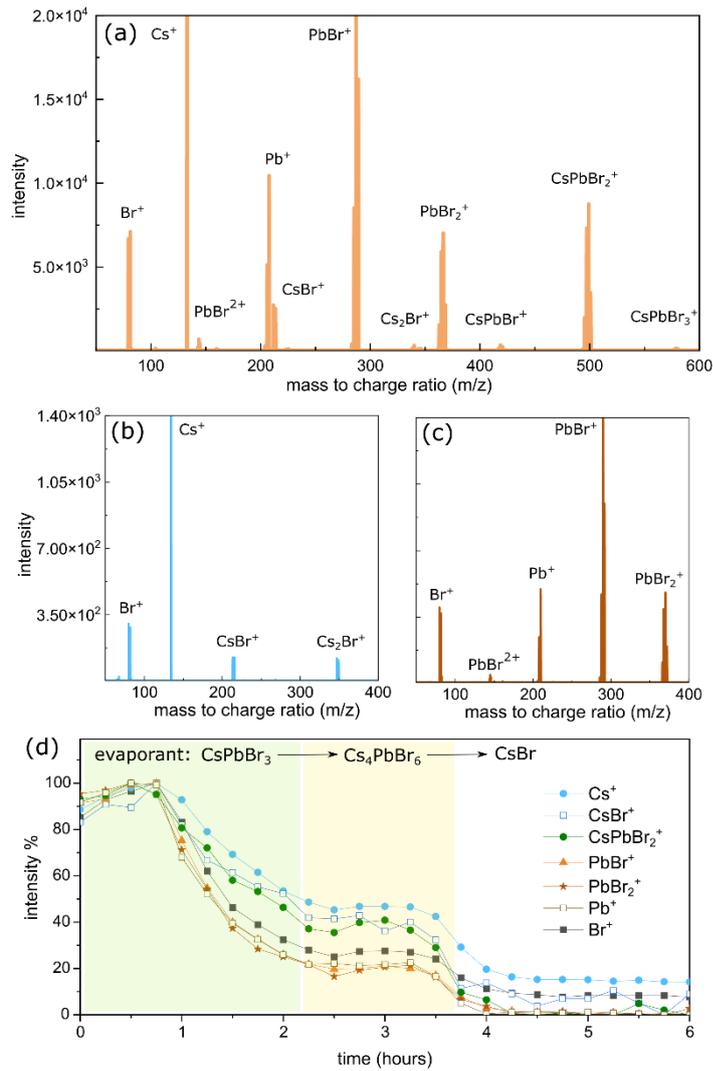

Fig. 2: KEMS results: mass spectra of (a) CsPbBr$_3$ pellet kept at 450 °C, (b) CsBr powder at 450 °C, (c) PbBr$_2$ powder at 300 °C. (d) Isothermal evaporation at 450°C of the CsPbBr$_3$ evaporant, the most relevant components monitored in time (only selected components are shown for clarity; more data are in Supporting Information, Fig. S3). The PbBr-related components are depleted faster (faster decrease in the graph for PbBr$^+$, PbBr$_2^+$, Pb$^+$ and Br$^+$) and at a certain moment a phase change to Cs$_4$PbBr$_6$ occurs. After all PbBr$_2$ evaporates, only CsBr remains.

As we show later, a precise flux tuning is not the key ingredient in the growth of phase pure perovskites. Instead, one first needs to know what basic building blocks are evaporated from the source materials. We have utilized Knudsen effusion mass spectrometry (KEMS) to obtain this information for the single source (CsPbBr$_3$ pellet)- as well as dual source-evaporation

(CsBr and PbBr$_2$ powders). Fig. 2 shows a mass spectrum recorded for each of these source materials heated to the temperatures further used in this study: CsPbBr$_3$ at 450 °C (Fig. 2a), CsBr at 450 °C (Fig. 2b), and PbBr$_2$ at 300 °C (Fig. 2c). As expected, the CsPbBr$_3$ KEMS spectrum contains ions that do not appear in the spectra of CsBr or PbBr$_2$ alone. Most importantly, CsPbBr$_2^+$ (493-503 m/z) which results from the fragmentation of CsPbBr$_3$ molecules after ionization. Although the fragmentation is very efficient (compare the intensity of CsPbBr$_2^+$ and CsPbBr$_3^+$ peaks in Fig. 2a), the presence of these components in the mass spectrum is indisputable proof that a congruent evaporation of CsPbBr$_3$ occurs as well. However, the congruent evaporation is only a partial, although the major process. This is documented by the existence of Cs$_2$Br$^+$, which can hardly originate from the CsPbBr$_3$ molecular clusters alone. Its presence indicates that (CsBr)$_x$ clusters evaporate as well (x≥2)[39]. Therefore, in addition to the congruent evaporation, the precursor in the crucible partially decomposes into CsBr and PbBr$_2$, and these components evaporate (at different rates, see below) simultaneously with the congruently evaporated CsPbBr$_3$ (see Fig. 2d). An extended analysis of KEMS data, including detailed temperature dependencies that further support these conclusions, is shown in the supporting information (Fig. S4, S5 and S6). The equilibrium vapor pressure of PbBr$_2$ is higher as compared to CsBr; therefore, if kept at the same temperature, PbBr$_2$ evaporates faster (compare the absolute ion currents in Fig. 2b,c). This fact is also documented in Fig. 2d, where the isothermal evaporation of CsPbBr$_3$ is monitored over time. The intensity of PbBr$_2$-related ions decreases faster compared to CsBr-ones. Therefore, the composition of the evaporant in the crucible gradually changes over time towards the Cs-rich phase. Consistent with the CsBr-PbBr$_2$ phase diagram[34], when a certain stoichiometry is reached (ratio 4:1), the phase change of the evaporant to Cs$_4$PbBr$_6$ occurs. This phase change is accompanied by different evaporation rates, followed by a rapid loss of PbBr-related components (clearly distinguishable in Fig. 2d).

Then, only CsBr is left within the crucible. Thus, KEMS measurements explain the previous experimental findings reporting different evaporant compositions after growth runs[31,33].

The congruent evaporation typically occurs in nanocrystalline evaporants because the heat is delivered abruptly to the entire nanocrystal, leading to a rapid decomposition and evaporation. Therefore, the preparation method is expected to be important. Nevertheless, we have found that different forms of the evaporant produce very similar results (see Supporting Information, Figs. S4). Additionally, for evaporant temperatures in between 400 °C and 450 °C, the vapor composition (ratios between components) is stable with the temperature (see Fig. S6). Thus, if operated before the phase change to $Cs_4PbBr_6$ and taking into account that the vapor pressure depends on temperature, the single-source $CsPbBr_3$ evaporation also allows to control the deposition rate via changing the evaporant temperature.

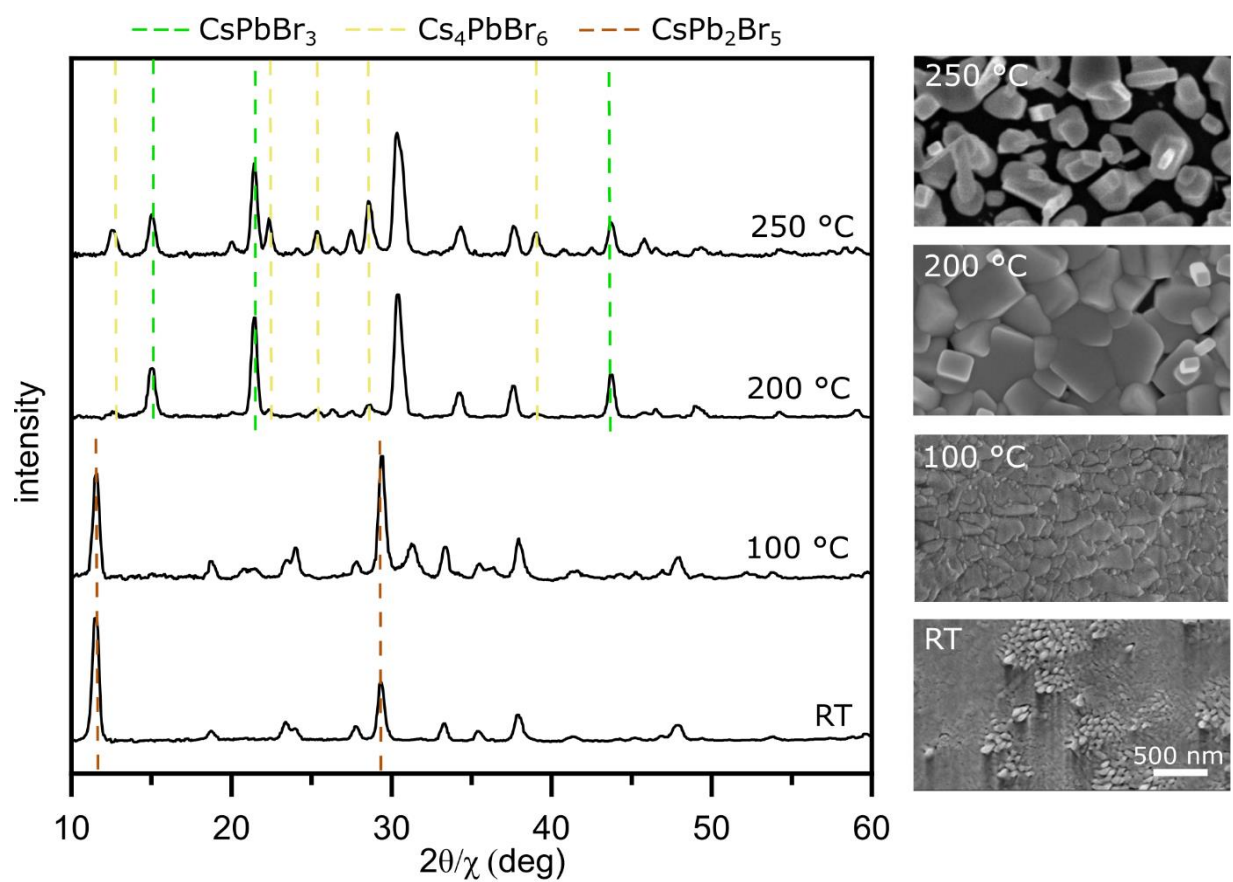

Fig. 3: Analysis of single source ($CsPbBr_3$, 450 °C) deposited perovskite layers at different substrate temperatures (RT, 100 °C, 200 °C, 250 °C) held for 30 minutes. (a) XRD analysis - significant peak

positions, unique to each stoichiometry, are marked by dashed color lines (see Fig. S1 for the full spektra and full peak assignments). (b) Representative SEM images of each sample.

Next, we show that it is possible to tune (to some extent) the stoichiometry of the deposited layers by controlling the substrate temperature without an extensive flux control. First, in Fig. 3 we show the XRD and SEM analyses of layers deposited from a single source evaporant ($CsPbBr_3$) at substrate temperatures ranging from room temperature (RT) to 250 °C. All growth runs were performed using a fresh evaporant at 450 °C to avoid changes in flux compositions due to phase changes in the evaporant (see Fig. 2d). Both the RT and 100 °C depositions yield a compact layer of $CsPb_2Br_5$. Upon increasing the substrate temperature, the grain size of the deposit slightly increases, while the nominal layer thickness corresponds to the deposition rate (470 nm). Conversely, we find that the morphology of the layer deposited at 200 °C is strikingly different. The individual grains possess a nearly-cubic shape and, occasionally, nanowires of a rectangular cross-section (up to 300 nm size) and maximum length 2 μm are protruding the polycrystalline layer. We argue that the appearance of the layer is clearly related to the stoichiometry change. At 200 °C, exclusively the $CsPbBr_3$ phase is identified in XRD spektra (see left panel in Fig. 3), identifying only a very minor presence of $Cs_4PbBr_6$. At a higher sample temperature (250 °C), the fraction of $Cs_4PbBr_6$ slightly increases. At even higher temperature (300 °C, not shown), no layer is formed, as all deposited material is instantly desorbed. Further analyses by XPS and Raman spectroscopy (see Supporting Information, Fig. S7 and S8) support the stoichiometry identified via XRD.

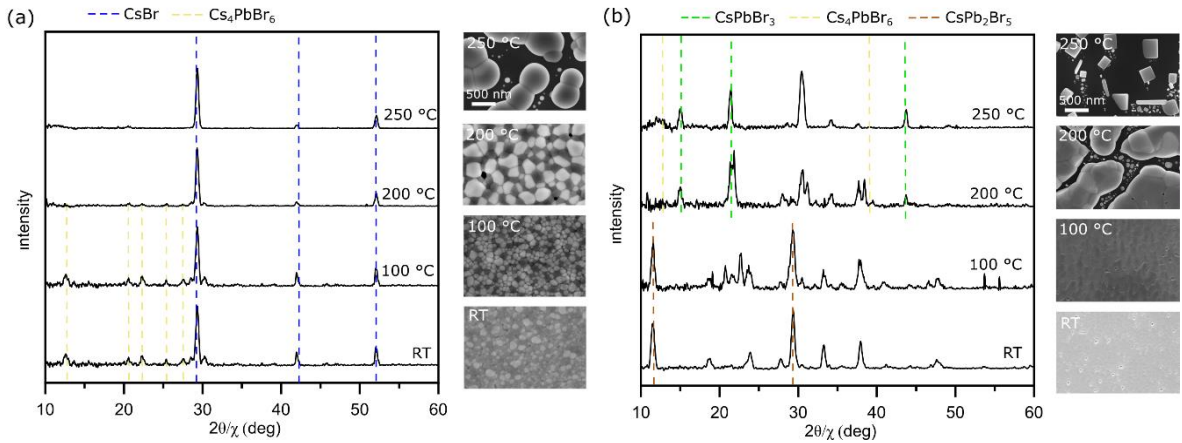

Fig. 4: Analysis of dual source (CsBr and PbBr$_2$) deposited perovskite layers at different sample temperatures (RT, 100 °C, 200 °C, 250 °C, 300 °C) and flux ratios. (a) Shows XRD and SEM analysis of the layers deposited with a PbBr$_2$/CsBr flux ratio of 1:5. The characteristic X-ray peak positions, unique to each stoichiometry, are marked by the vertical dashed colored lines (see Fig. S1 for the full spectra). In (b) the same analysis is made for a PbBr$_2$/CsBr flux ratio of 3:1.

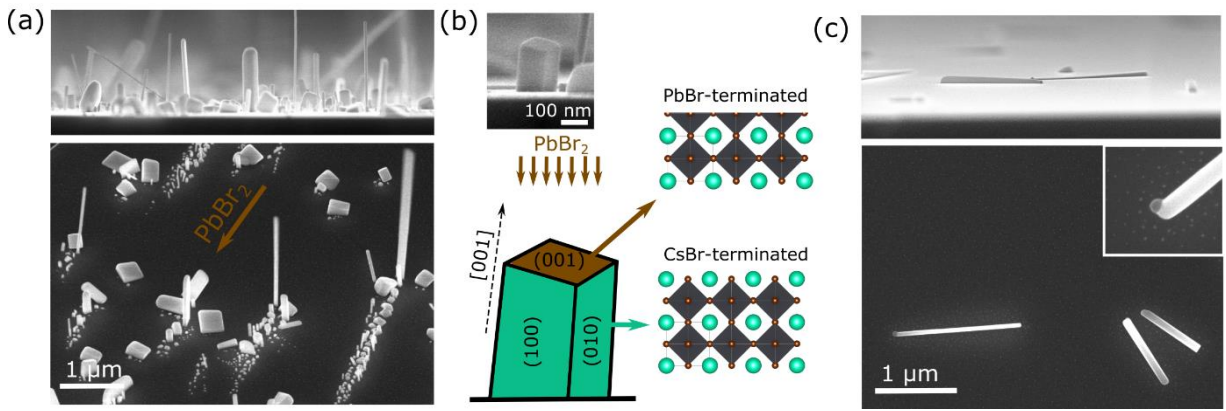

Fig. 5: CsPbBr$_3$ nanowires observed when higher substrate temperatures and PbBr$_2$ flux (dual source deposition) were used. In (a), a detailed SEM inspection of the sample grown at 250 °C (same as in Fig. 4b, PbBr$_2$/CsBr flux ratio of 3:1) is shown, with the side view in the top row, and a tilted (45°) view in the bottom row. At 250 °C, out-of-plane CsPbBr$_3$ nanowires are formed, with cuboid CsPbBr$_3$ crystals mostly within the shade of nanowires in the direction of the PbBr$_2$ beam (the beam direction is marked by an arrow, the sample was not rotated during the evaporation). (b) A SEM image of a slightly elongated cuboid crystal, with a clearly visible square top facet, which is identical to the nanowire cross-section. The schematic illustration shows the proposed growth model for anisotropic growth. (c) The sample grown at 300 °C using a very high PbBr2/CsBr flux ratio of 25:1, where exclusively in-plane oriented CsPbBr$_3$ nanowires are formed. The inset shows a detail of a nanowire tip with a nanoparticle, typical for vapor-solid-solid growth.

The dual source deposition (simultaneous independent evaporation of CsBr and PbBr$_2$) allows tuning the PbBr$_2$/CsBr flux ratio. Fig. 4a shows XRD and SEM analysis of layers deposited at sample temperatures ranging from RT to 250 °C when the flux ratio is smaller than 1 (in this case, the PbBr$_2$/CsBr flux ratio was 1:5). All the layers are composed mostly of CsBr. Only for lower sample temperatures (RT and 100 °C), a certain fraction formed by Cs$_4$PbBr$_6$ is detected. As expected, the grain size increases with temperature. At 250 °C, the layer is partially decomposed and roughened (similarly to the single-source deposition). When the PbBr$_2$/CsBr flux ratio is changed to 3:1 (Fig. 4b), the results are mostly similar to those of the single-source evaporation. The compact layer of CsPb$_2$Br$_5$ is deposited at low temperatures (RT and 100 °C), while at higher temperatures, CsPbBr$_3$ is formed. However, in contrast to the single source evaporation, there are several noticeable differences. Firstly, as revealed by XRD, pure CsPbBr$_3$ is formed at 250 °C. At this temperature, the Cs$_4$PbBr$_6$ present in samples prepared by the single source evaporation is absent. Secondly, CsPbBr$_3$ forms well-defined rectangular and cuboid crystals as well as nanowires.

We have inspected the sample grown at 250 °C in more detail (see Fig. 5a). The nanowires on the sample were all out-of-plane, no in-plane nanowires were observed. All the nanowires have a square cross section (see detail in Fig. 5b). Interestingly, cuboid crystals present on the sample as mostly formed in the proximity of nanowires; almost exclusively in the shade of nanowires in the direction of the PbBr$_2$ flux. The largest difference between the single- and dual-source evaporation is that using the latter approach the growth is possible also at temperatures above 250 °C. In order to do so, the PbBr$_2$/CsBr ratio has to be much higher compared to the one achievable by the single source deposition. Fig. 5c shows that at 300 °C and under the very high PbBr$_2$ flux (PbBr$_2$/CsBr flux ratio of 25:1), only small crystallites and relatively long CsPbBr$_3$ nanowires are present on the surface. Interestingly, in contrast to the deposition at 250 °C, all the nanowires grow in-plane, along the substrate surface. In many cases, a small droplet is visible at the nanowire end (see the inset in the bottom panel of Fig. 5c), and the nanowires frequently exhibit a tapered morphology.

## IV. Discussion

Comparing the single- and dual-source deposition results, one comes to the following conclusions. At low deposition temperatures, our results agree with those reported by previous studies[32,33], that is both deposition approaches result in the $CsPb_2Br_5$ phase, occasionally mixed with $CsPbBr_3$. In order to grow the pure $CsPbBr_3$ phase, one needs to raise the sample temperature. The KEMS data, together with the analyses shown in Fig. 3 and Fig. 4 allow us to explain the mechanism behind. An elevated sample temperature results in desorption of excess $PbBr_2$ from the sample surface. The desorption of $PbBr_2$ prevents the formation of the $CsPb_2Br_5$ phase at temperatures above 100 °C. Instead, at elevated sample temperatures the growth rate becomes limited by the CsBr flux. This makes the $CsPbBr_3$ growth mechanism on surfaces very similar to that of III-V semiconductors formation via molecular beam epitaxy: in excess of group V species, the growth is controlled by the group III element. Here, the large flux of $PbBr_2$ allows the stoichiometric formation of $CsPbBr_3$, while the elevated sample temperature ensures the desorption of excess $PbBr_2$. This is valid for both the single-source and dual-source (with $PbBr_2$/CsBr flux ratios >1) depositions. The $PbBr_2$/CsBr flux ratio should be large; otherwise, the desorption of $PbBr_2$ could be so fast that growth of $CsPbBr_3$ can become $PbBr_2$-limited. This is the case of single-source deposition at 250 °C (Fig. 3). In such a case, the lack of $PbBr_2$ (and, hence, excess of CsBr) results in a partial formation of $Cs_4PbBr_6$. The desorption of $PbBr_2$ is a temperature-activated proces; hence, it can be compensated by a very large $PbBr_2$ flux. This is possible by the dual-source deposition only and, thus, $CsPbBr_3$ can be grown even at sample temperatures above 250 °C. As shown in Fig. 4 and 5, as a consequence of the elevated temperature and high $PbBr_2$ flux, the $CsPbBr_3$ deposit morphology rapidly changes from a fairly compact layer to nanowires. The appearance of nanowires during the vapor phase deposition of inorganic perovskites at elevated temperatures has been observed previously[17,22,40]. This morphology change is still unexplored, despite attempts to explain it by temperature-dependent changes in the crystal structure[11]. Here, we observe two different cases. At the highest sample temperature of 300 °C inspected in this study (which is still below the liquidus line of the CsBr-PbBr$_2$ phase diagram), Cs-Pb-Br alloy nanoparticles are formed that further collect the deposited species, facilitating one-dimensional growth via vapor-solid-solid (VSS) mechanism[41]. The appearance of nanoparticles at the tip of the nanowires is a typical signature of VSS growth. The in-plane nanowire geometry reached at this deposition temperature results from a nanoparticle movement on the surface during growth[42]. The nanoparticles shrink in time (and, potentially, fully diminish after some time), which is reflected in the tapered shape of the in-plane nanowires (Fig.5c, side view).

At the slightly lower sample temperature of 250 °C, but still under high PbBr$_2$ flux, the out-of plane nanowires do not exhibit any nanoparticle at their tip. As the nanowire cross-section does not change along their axis, the absence of the droplet cannot be explained by its evaporation during growth. There is obviously another growth mechanism in play. We propose that the asymmetry promoting one-dimensional growth in the system is the distinct surface termination of CsPbBr$_3$ facets. Cubic CsPbBr$_3$ crystals, as is the case here (Fig. 3), commonly exhibit (100) facets[17]. These facets are crystallographically identical, and under common growth conditions they are Cs-Br terminated due to the lowest surface free energy of this termination[43,44,45]. We hypothesize that, at PbBr$_2$-rich conditions, the facet that is most directly exposed to the incident flux converts into a Pb-Br terminated one. Such a facet immediately develops to the fastest growing one due to an increase in the surface free energy[43]. As a result, cubic crystals elongate along the [100] direction (Fig. 5b) and form nanowires with a square cross-section, as observed experimentally (Fig. 5a). The elevated growth temperatures promote the relevant kinetic processes (mostly surface diffusion), further accentuating the anisotropy of the growth.

**Conclusions**

In summary, we have demonstrated how differences in the vapor pressures of CsBr and $PbBr_2$ affect the vapor phase growth from both single- ($CsPbBr_3$) and dual-sources (CsBr and $PbBr_2$). Under an excess $PbBr_2$ flux, the growth of $CsPbBr_3$ is CsBr-limited. Hence, if growth is conducted at certain elevated sample temperatures, this mechanism allows full control of the growth rate by the CsBr flux in case of the dual source depositon. Similarly, it allows the deposition of the pure $CsPbBr_3$ phase. Our study shows that both single- and dual-source depositions of inorganic perovskites may give similar results if appropriately controlled. The single-source deposition, however, requires caution; longer operation results in a phase change of the evaporant and related flux variations. Additionally, the dual-source deposition is more flexible, because it allows tuning the fluxes of each component independently. Flux variations broaden the variety of $CsPbBr_3$ morphologies, ranging from cubic nanoparticles to nanowires. We have proposed a nanowire mechanism specific for $CsPbBr_3$ that explains the nanowire growth at high temperatures via changes in surface-termination of the top facet under an excessive $PbBr_2$ flux.

**Acknowledgements**

This article is partially based upon work from COST Action OPERA – European Network for Innovative and Advanced Epitaxy – CA20116, supported by COST (European Cooperation in Science and Technology, www.cost.eu). CzechNanoLab project LM2023051 funded by MEYS CR is gratefully acknowledged for the financial support of the measurements at CEITEC Nano Research Infrastructure. This work was supported by the project Quantum materials for applications in sustainable technologies (QM4ST), funded as project No. CZ.02.01.01/00/22_008/0004572 by OP JAK, call Excellent Research, and the authors also acknowledge funding by Specific Research of BUT FCH/FSI-J-24-8514.

# Achieving Different Stoichiometries and Morphologies in Vapor Phase Deposition of Inorganic Halide Perovskites: Single or Dual Precursor Sources?

Tomáš Musálek[1,2], Petr Liška[1,2], Amedeo Morsa[3], Jon Ander Arregi[2], Matouš Kratochvíl[4], Dmitry Sergeev[3,5], Michael Müller[3], Tomáš Šikola[1,2], Miroslav Kolíbal[1,2,*]

[1]Institute of Physical Engineering, Brno University of Technology, Technická 2, 616 69 Brno, Czechia

[2]CEITEC BUT, Brno University of Technology, Technická 10, 61669 Brno, Czechia

[3]Institute of Energy Materials and Devices, Structure and Properties of Materials (IMD-1), Forschungszentrum Jülich, 52425 Jülich, Germany

[4]Materials Research Centre, Faculty of Chemistry, Brno University of Technology, Purkyňova 118, 612 00 Brno, Czech Republic

[5]NETZSCH-Gerätebau GmbH, Selb, D-95100, Germany

* mail: kolibal.m@fme.vutbr.cz

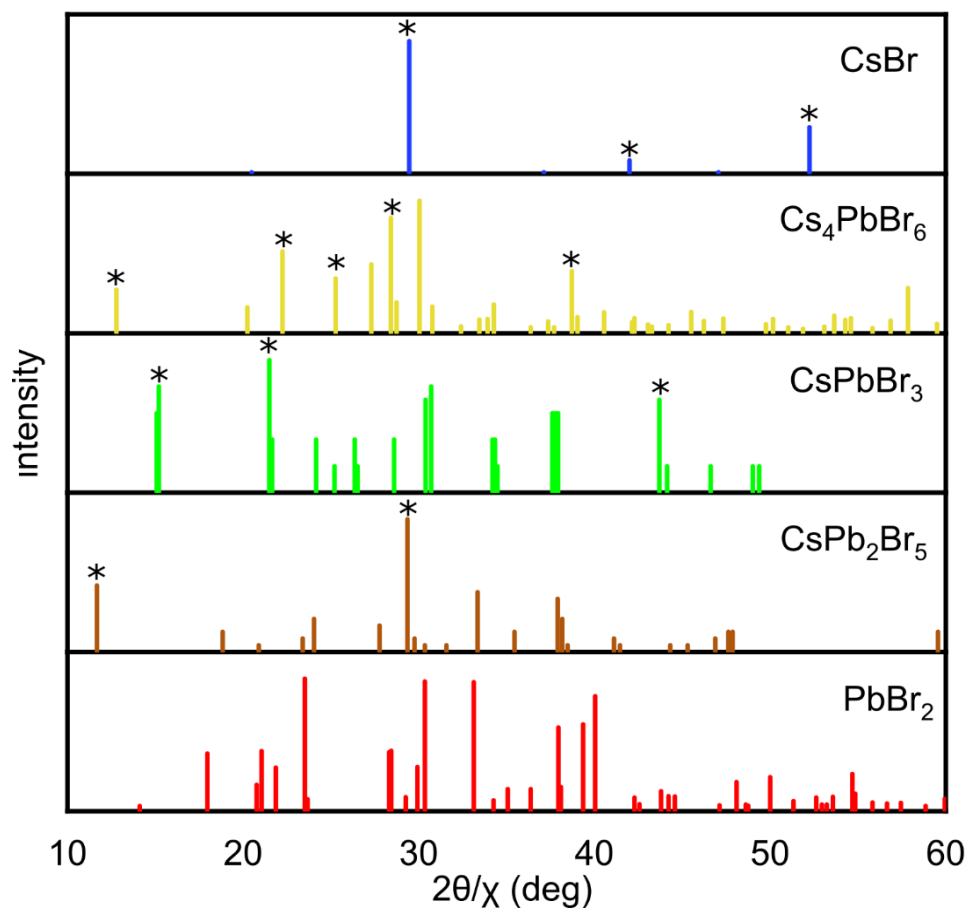

Fig. S1: XRD data taken from [CsBr - PDF Card No.: 00-001-0843, CsPb$_2$Br$_5$ - PDF Card No.: 00-025-0211, CsPbBr$_3$ - PDF Card No.: 00-018-0364, Cs$_4$PbBr$_6$ - materialsproject.org, PbBr$_2$ - materialsproject.org]. The most significant peaks, which were used for identification of each phase in Figures 3 and 4, are marked with asterisks.

**Knudsen mass spectrometry experimental details**. The FINNIGAN MAT 271 spectrometer is composed of four main components: a Knudsen cell (KC), an electron impact ion source, a single-focusing magnetic sector-field mass spectrometer, and a detection system consisting of a secondary electron multiplier with an ion counter.[S1] The sample to be analysed is placed inside an alumina liner with a height of 8 mm, an outer diameter of 6.4 mm and a wall thickness of 0.5 mm, which in turn is placed in an iridium Knudsen cell consisting of a crucible measuring 8.5 mm in height, with an outer diameter of 7.8 mm and a wall thickness of 0.2 mm. The cell is covered with a lid that has a central orifice of 0.4 mm, allowing for the effusion of the molecular beam. To maintain the desired temperature conditions, the KC is insulated by three nested heat shields made of tantalum. The tantalum shields are designed with openings to facilitate the effusion of a molecular beam along the cell's axis. Heating of the KC is achieved through two methods: radiation heating is used to reach temperatures of approximately 873 K, while higher temperatures are achieved via electron bombardment, in both cases by means a tungsten heating wire twisted around the KC.[S1,S2] A $W_{97}Re_3/W_{75}Re_{25}$ thermocouple, integrated into the Knudsen cell holder, monitors the temperature of the chamber, while the temperature at the KC is measured by a pyrometer (Dr. Geaorg MAURER GmbH -Optoelektronik-) through a hole in the tungsten sample-holder beneath the KC (Fig. S2, inset). The entire setup is contained within a vacuum chamber maintained at approximately $10^{-6}$ mbar by a turbo molecular pump, and isolated from the mass spectrometer compartments, which are kept under ultrahigh vacuum (~$10^{-9}$ mbar) by ion getter pumps. The setup includes a shutter mechanism that allows for distinguishing between sample signals and background noise by blocking or allowing the molecular beam to pass. When the shutter is open, the molecular beam effuses from the KC through an aperture into the ion source, where ionization of the gaseous species occurs. Ionization is achieved via an electron beam generated by an tungsten cathode, operating at an electron energy between 60 and 70 eV with an emission current of 0.47 mA. The detection system comprises an secondary electron multiplier and an ion counter. The ion-counting method, which reduces measurement errors through mass discrimination by the multiplier, was employed in the experiments conductedin this study.

Before introducing samples, the Knudsen cells were preconditioned by heating at 1000 °C for 12 hours to eliminate impurities. About 40 mg of powder under study was loaded into the cell. The cell, along with its contents, was weighed. Upon identifying a sufficiently intense signal the KC's position was adjusted to optimize the signal. Two types of measurement series were conducted: isothermal and polythermal. The isothermal series involve measuring vapor species at a fixed temperature over a specified period, while the polythermal series includes measurements taken at different temperatures.

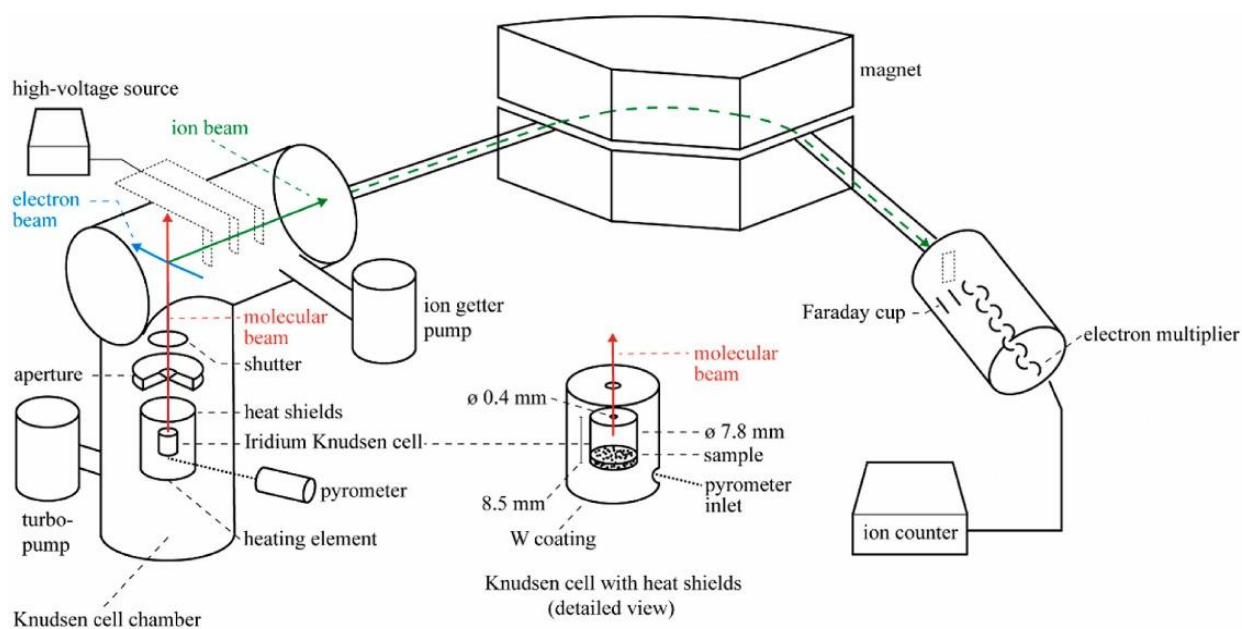

Fig. S2: Schematic view of a magnetic-sector Knudsen effusion mass spectrometer (for detailed description, see the main text). Adopted from Ref. S3.

**Additional KEMS data.**

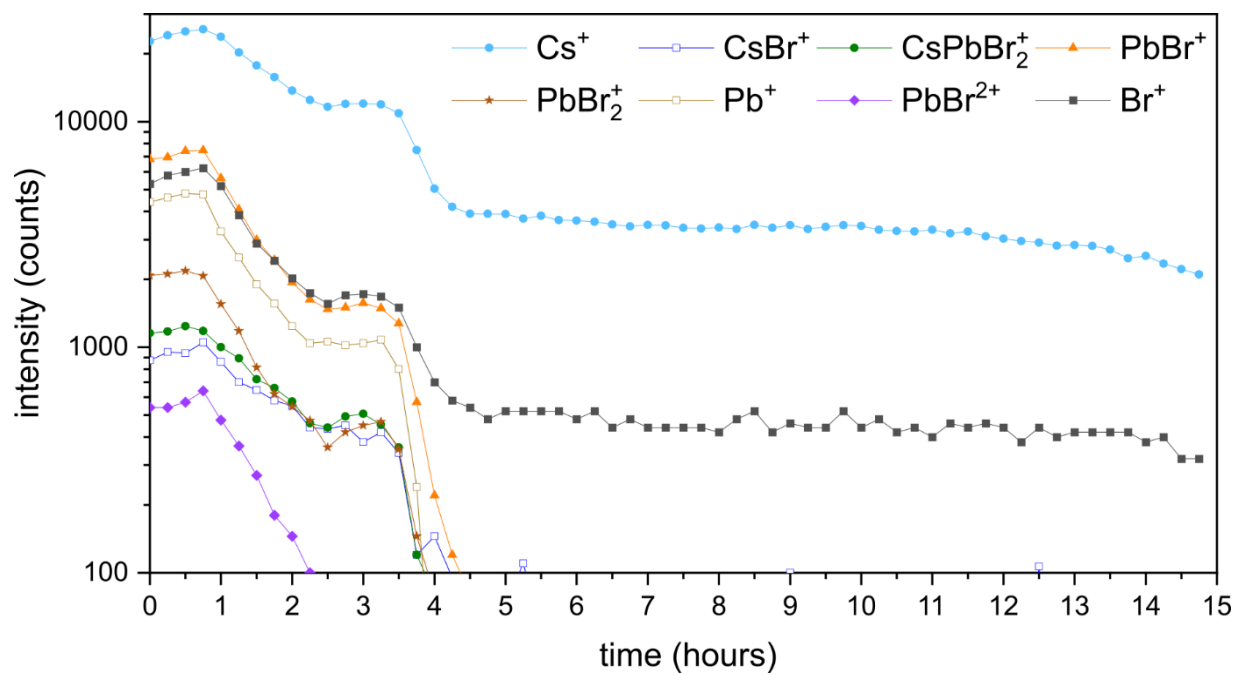

Fig. S3: A full monitored spectrum of the m/z signals obtained during isothermal evaporation at 450 °C of the CsPbBr$_3$ evaporant. In comparison to Fig. 2d, the full time scale is shown, as well as an additional signal from PbBr$^{2+}$.

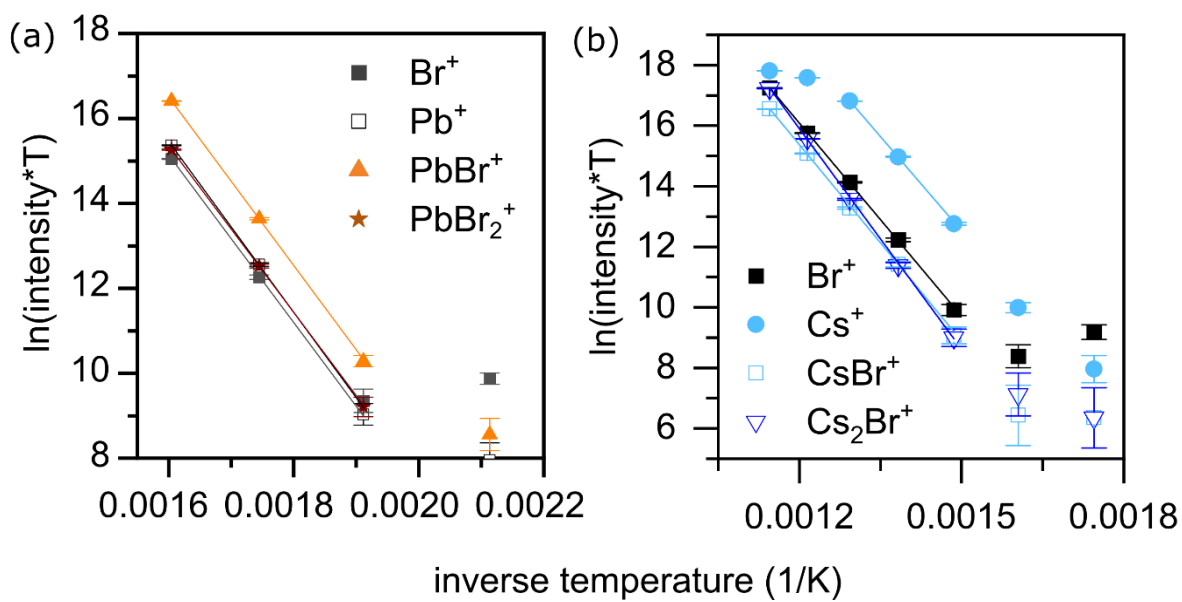

Fig. S4: Clausius-Clapeyron analysis helps to understand the nature of the molecule fragmentation during KEMS measurements. It allows to discriminate between different evaporated clusters and fragments of the same cluster, emergent during the ionization process in the spectrometer. The depicted dependence of the logarithm of the peak intensity multiplied by temperature on inverse temperature is linear. Fragments of the same cluster share the same slope; on the contrary, different clusters (and their fragments) should exhibit a distinct slope. All the ions in (a) exhibit the same slope (within the measurement uncertainty), clearly identifying them as fragments of the same cluster $(PbBr_2)_1$, being in accordance with the literature.[S4] In (b), CsBr evaporation analysis gives an indication of other clusters than $(CsBr)_1$ only. It is plausible to assume also other clusters to evaporate from CsBr, namely $(CsBr)_x$, with $x \geq 1$.[S5] The $Cs_2Br^+$ is a product of the fragmentation of the $(CsBr)_2$ dimer; as a confirmation, the relevant slope in (b) differs from that of $CsBr^+$ (the difference is larger as compared to the measurement uncertainty). The data at low temperatures deviate from the linear dependence due to very low counts in the spectra; hence, they were not considered relevant during the linear fitting. The uncertainties were calculated as a square root of each peak count.

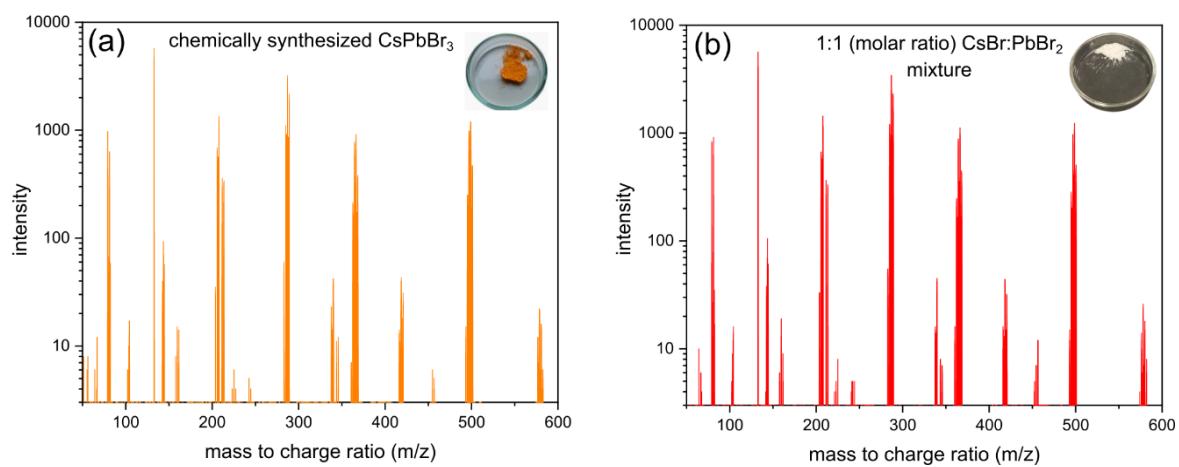

Fig. S5: Mass spectrum of (a) a chemically synthesized $CsPbBr_3$ compared to (b) the mass spectrum of two powders (CsBr and $PbBr_2$) mixed in the molar ratio 1:1, after 2 hours heating at 400 °C prior to the mass spectrometry. Both spectra were taken at a precursor temperature of 400 °C and are almost identical, despite the initial visual difference (photographs of the two precursors are in the insets). Clearly, inevitable annealing of the powder mixture in (b) in vacuum results in the formation of $CsPbBr_3$ inside the crucible after some initial period. After that, the mass spectrum is the same as the initially synthesized $CsPbBr_3$.

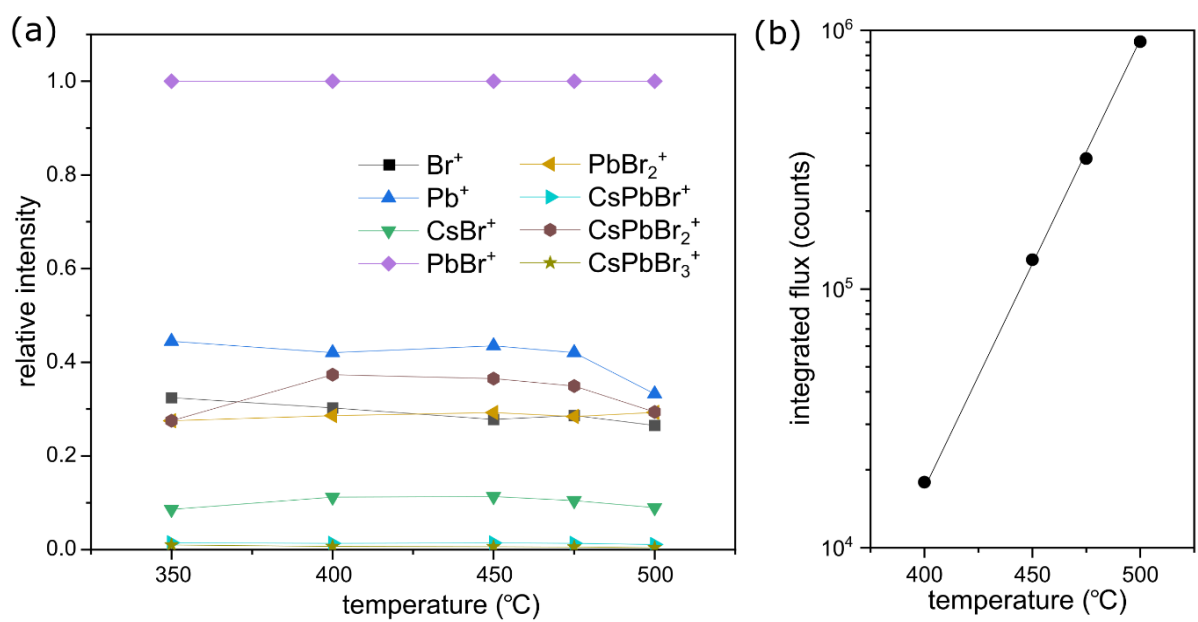

Fig. S6: (a) Peak intensities extracted from KEMS spectra of the CsPbBr$_3$ evaporant recorded at different temperatures, normalized to the PbBr$^+$ component. A stable ratio of the relevant components is achieved within a temperature range of 400-450 °C. The sum of the peak intensities of critical components (CsBr$^+$, PbBr$_2^+$ and CsPbBr$_3^+$, excluding fragments) is shown in (b); it is proportional to the flux towards the sample. Hence, within the mentioned temperature range, a control of the vapor flux (with a fixed vapor composition) during the single source evaporation is possible by changing the evaporant temperature.

**XPS analysis of the deposits**. Identification of different perovskite stoichiometries based solely on binding energies of each constituent is an extremely challenging task. However, we have found many correlations with XRD results, which enable the identification and discrimination between CsPbBr$_3$, Cs$_4$PbBr$_6$, and CsPb$_2$Br$_5$ stoichiometries.

As already noticed in Fig. 1, specific care has to be taken with respect to the X-ray dose delivered to the sample during the measurement. This is already a well-known problem, where XPS helped to determine the mechanism – X-Ray exposure causes dissociation and clustering of Pb atoms, which are detected as the metallic Pb$^0$ component in the spectrum. In agreement with previous literature,[S6] we have found that CsPbBr$_3$ is very resistant to degradation under X-rays and the sample did not exhibit the detectable Pb$^0$ component after the routine measurement (1486 eV X-Ray fluence of $5\times10^{16}$ photons/cm$^2$). A presence of the Pb$^0$ peak is thus a fingerprint of another phase, and XPS sampling of this component can be utilized as a sensitive measure of phase purity. Our results indicate that $Cs_4PbBr_6$ is the most susceptible to an X-ray degradation, while $CsPb_2Br_5$ showed a mediocre surface damage.

Although many authors attempted to identify the phases by the binding energy shifts of specific peak components, this approach is inconclusive and could even be misleading. Different Fermi level shifts and sample charging usually require the use of e.g. neutralizer. In such a case, the spectra have to be shifted during processing, which introduces significant uncertainty and complicates the correct interpretation. An analysis of core-to-core level shifts overcomes these issues and allows to identify specific fingerprints for two of the studied phases. $CsPbBr_3$ can be identified by the difference in a core-to-core level between Pb $4f_{5/2}$ and Br $3d_{5/2}$, and Pb $4f_{5/2}$ and Cs $4d_{5/2}$, being specifically (70.00-70.15) eV and (62.88-68.92) eV, respectively (see Fig. S7). Also, we have noticed that the Pb $4f_{5/2}$ peak width is wider than for other stoichiometries; for our experimental settings it varies between 1.1 and 1.4 eV, while other perovskites always show full widths-half-maxima below 1 eV. The origin of this peak widening has not yet been determined.

$CsPb_2Br_5$ can be identified from differences between Cs $3d_{5/2}$ and Br $3d_{5/2}$, and Cs $4d_{5/2}$-Br $3d_{5/2}$, being (655.50-655.65) eV and (6.50-6.65) eV, respectively (see Fig. S7). No clear fingerprints of $Cs_4PbBr_6$ have been identified, as the binding energies strongly resemble those of pure CsBr.

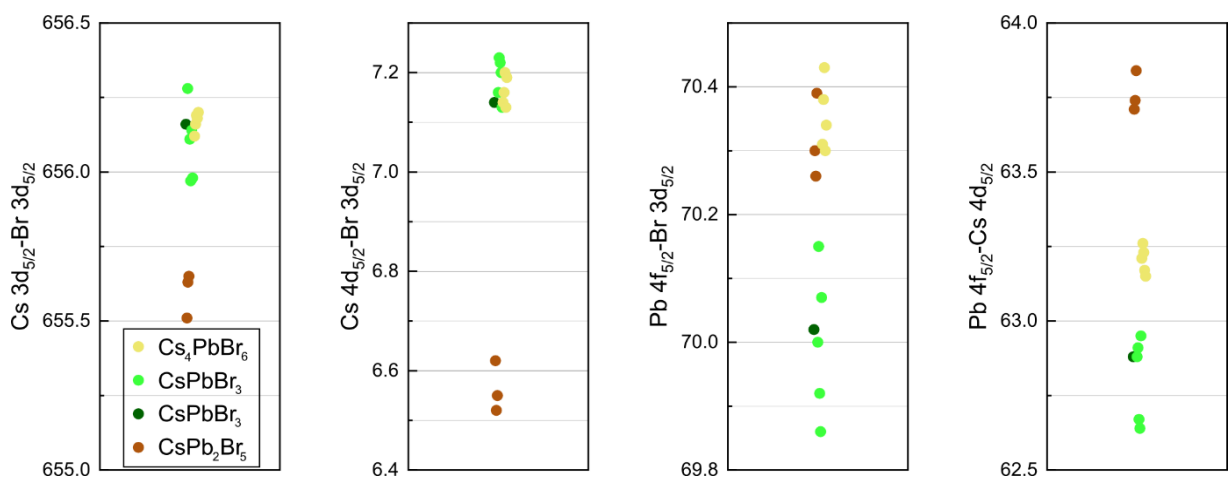

Fig. S7: XPS core-to-core level shifts. These were calculated for all samples that exhibited one of the phases as the major phase according to the XRD spectra. The dark green spots mark the bulk $CsPbBr_3$ for comparison. Clearly, there are correlations to be found for $CsPbBr_3$ and $CsPb_2Br_5$. A scatter of the data (specifically, for example, Pb $4f_{5/2}$ – Br $3d_{5/2}$ for $CsPbBr_3$) is caused by the fact that not all samples were phase pure; minor phases present in the samples weaken the core level shift differences.

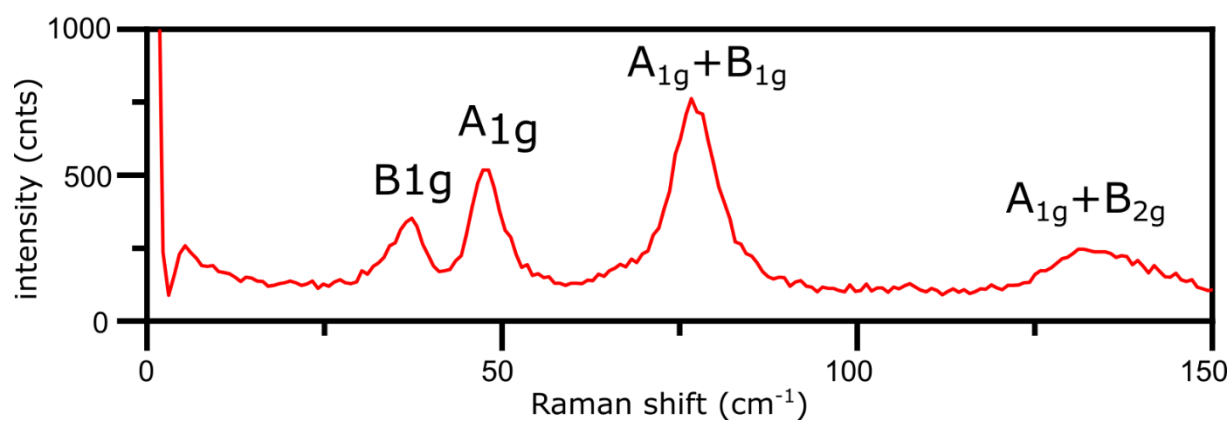

Fig. S8: Raman spectrum taken from a single-source deposited sample at 100 °C (XRD and SEM analysis in Fig. 3). The peaks are a fingerprint of the $CsPb_2Br_5$ phase.[S7]